\pdfoutput=1
\documentclass{article}
\usepackage{spconf,amsmath,graphicx,hyperref,booktabs,multirow,xcolor,kotex}
\usepackage{orcidlink}  

\newcommand{\method}{STEMMA}
\newcommand{\bench}{STEMMA-Bench}
\newcommand{\train}{STEMMA-Instruct}

\title{STEMMA: SONG-TO-STEM MULTI-AUDIO REASONING FOR LARGE AUDIO LANGUAGE MODELS}

\name{\shortstack[c]{
Hoyeol Sohn$^{1}$\hspace{1pt}\orcidlink{0009-0004-5608-7018}\thanks{This work was conducted while the first author was an intern at Neutune.},
Wonil Kim$^{2}$\hspace{1pt}\orcidlink{0000-0002-6105-0488},
Keunhyoung Kim$^{2}$\hspace{1pt}\orcidlink{0000-0001-8167-5752},
Sangeun Kum$^{2}$\hspace{1pt}\orcidlink{0000-0001-9305-7281},
Taehyoung Kim$^{2}$\hspace{1pt}\orcidlink{0009-0005-1427-2976}, \\
Dongjoo Moon$^{2}$\hspace{1pt}\orcidlink{0009-0000-8998-3629},
Theerasak Charoenchob$^{2}$\hspace{1pt}\orcidlink{0009-0001-9416-131X},
Teeratep Weerapang$^{2}$\hspace{1pt}\orcidlink{0009-0009-7697-1472},
Jongpil Lee$^{2}$\hspace{1pt}\orcidlink{0000-0002-1126-0081},
Juhan Nam$^{1}$\hspace{1pt}\orcidlink{0000-0003-2664-2119}}
}

\address{
$^{1}$ Graduate School of Culture Technology, KAIST, Daejeon, Republic of Korea \\
$^{2}$ Neutune, Seoul, Republic of Korea
}

\begin{document}
\ninept
\maketitle

\begin{abstract}

Music understanding often requires comparing excerpts and reasoning about relationships among songs, sections, and stems. However, existing large audio-language models (LALMs) and music question-answering datasets typically operate on single recordings or compare independently sampled tracks with no known production relationship. We introduce \method{}, a multi-audio music question-answering framework built around production provenance: whether excerpts originate from the same track or section, and which stems belong to which mixtures. Because such relations are sparse under conventional audio-first sampling, \method{} adopts a relation-first construction strategy: it first specifies a target relation and then queries the catalog for excerpts that satisfy it and hard negatives that do not. Labels are determined directly from catalog provenance rather than generated by a language model from metadata. We build \bench{} for evaluation and a track-disjoint training set, \train{}. Fine-tuning two LALMs on \train{} improves multi-audio reasoning, with the largest gains on structural relations directly determined by the catalog, while preserving single-audio music understanding.

\end{abstract}

\begin{keywords}
multi-audio reasoning, music understanding, audio-language models, instruction tuning, hierarchical music structure
\end{keywords}

\section{Introduction}
\label{sec:intro}

A line of music-specialized large audio-language models (LALMs) has shown that instruction tuning on music-specific supervision improves captioning, question answering, and analysis of music theory, structure, and instrumentation~\cite{llark,musicflamingo}. Nevertheless, most models and training corpora still treat each recording as an independent item.

Emerging generative music experiences increasingly require models to reason about relationships among songs, sections, stems, and musical attributes~\cite{mulligan2026future}. %
Retrieval, arrangement, and sample discovery therefore require a model to determine whether excerpts originate from the same track or section~\cite{sampleid} and to compare tempo, key, instrumentation, harmony, genre, and mood~\cite{kim2025generation,stemretrieval}. This setting requires \emph{cross-audio grounding}: evidence must be localized in each input before a relation across inputs can be inferred.
Conversational music agents orchestrate MIR and generation tools from natural language but typically rely on intermediate tool outputs and stored musical attributes rather than jointly reasoning over multiple audio inputs~\cite{musicagent}. As a result, relations requiring direct comparison between signals remain outside their native reasoning scope.

Recent work has begun to address this limitation: explicit multi-audio instruction tuning improves comparison and reasoning~\cite{beyondsingleaudio,polyaudio}, while comparative question answering has been extended to pairs of independently collected music tracks~\cite{jamendomtqa} and to natural-language explanations of audio difference~\cite{adiff}. Performance nevertheless degrades as the candidate set grows~\cite{mugen}. However, prior work does not systematically exploit the known song--section--stem provenance available in music production data.

We propose \method, a multi-audio music question-answering framework whose answers are
determined by known song--section--stem provenance: whether two excerpts share a track or a section,
and which stem belongs to which mixture.
Because such relations are too sparse under conventional audio-first sampling, \method{} adopts a \emph{relation-first} strategy: the target relation is specified before audio is sampled, and a single catalog query retrieves both matching examples and near-misses that violate it.
Our contributions are: (1) \bench{}, a held-out
benchmark over sections and section stems drawn from full songs rather than over
independently sampled tracks, and (2) \train{}, a track-disjoint training set on
which we instruction-tune two LALMs, raising aggregate multi-audio accuracy
while single-audio music understanding is retained. Project page: \url{https://github.com/hoyso48/STEMMA}.

\begin{figure*}[t]
  \centering
  \includegraphics[width=\textwidth]{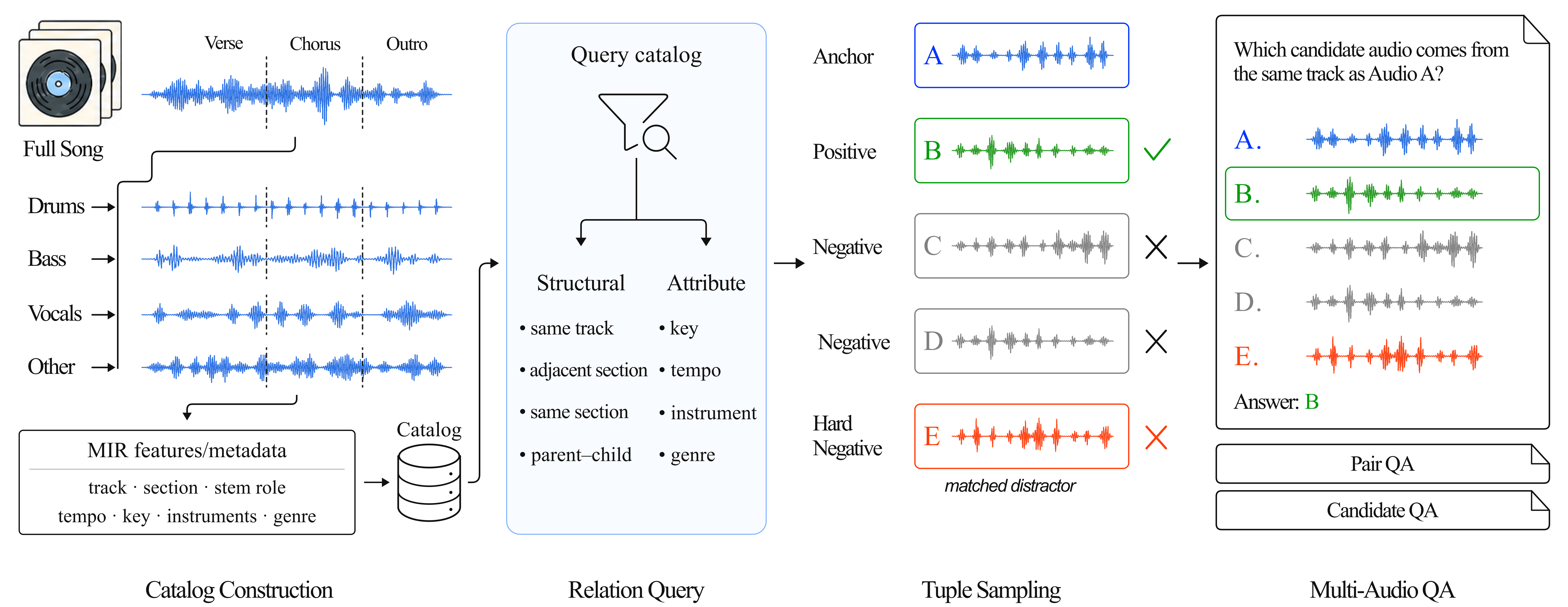}
  \caption{Overview of the proposed relation-first generation pipeline. Hierarchical music units are indexed by structural and musical-attribute relations. A relation is stated over catalog fields, and the tuples that satisfy or violate it are retrieved by query before being rendered as multi-audio instructions.}
  \label{fig:pipeline}
\end{figure*}

\section{Related Work}
\label{sec:related}

\textbf{Music understanding in LALMs.}
LLark unifies captioning, understanding, and reasoning over open-source music annotations in a single instruction-tuning format~\cite{llark}. Music Flamingo scales this approach with synthetic supervision covering harmony, structure, timbre, lyrics, and cultural context~\cite{musicflamingo}, while MOSS-Music is an open music audio-language model trained with a similar pipeline~\cite{mossmusic2026}. These models are trained and evaluated one recording at a time.

\textbf{Multi-audio grounding in LALMs.}
Beyond Single-Audio constructs paired-audio tasks and shows that targeted fine-tuning can add two-audio capability without substantially sacrificing single-audio performance~\cite{beyondsingleaudio}. PolyAudio extends this to interleaved contexts and four skill families, including comparison and temporal reasoning~\cite{polyaudio}. MUGEN evaluates grounding among multiple audio candidates, finding that performance falls with more inputs and that acoustic comparison is the main bottleneck~\cite{mugen}.

\textbf{Comparative music QA.}
Jamendo-MT-QA samples full-track pairs and uses an LLM to generate comparative questions from per-track captions, metadata, and single-track QA, followed by human- and LLM-based filtering~\cite{jamendomtqa}. Independently sampled tracks share no provenance, so the questions can only compare attributes. \method{} instead draws sections and stems from the same song, enabling structural questions that independent pairs cannot express.

\section{Relation-First Multi-Audio Data}
\label{sec:method}

\subsection{Why the relation comes first}

A relation-labeled tuple can either be sampled and retained when the relation happens to hold or queried directly. Structural relations are sparse. Indeed, two excerpts drawn independently from $N$ tracks share a track with probability $O(1/N)$, and sharing a section is rarer still, so sampling wastes most draws on rejected tuples, whereas direct querying eliminates this rejection cost by retrieving tuples that satisfy the relation.

Specifying the relation first also controls how similar the negative examples are to the correct one. This is defined by explicit conditions rather than by chance sampling. Because the answer is determined from catalog fields before any text is generated, neither the label nor the difficulty of the example depends on language-model output. A new task therefore only requires a new task definition and the query used to retrieve its examples, while reusing the same catalog and data format.

\subsection{Building the catalog}
\label{sec:catalog}

The first step is to turn music hierarchy into a queryable table (Fig.~\ref{fig:pipeline}).
Each track is divided into ordered sections with section labels, and each section is separated into individual musical roles. A single recording therefore yields many excerpts with known relationships.
Every excerpt corresponds to one section and contains either the full section mixture or one separated role.
We call these a \emph{section} and a \emph{section stem}, respectively. Each row stores four provenance fields: track, section index, section category, and stem role. It also stores musical attributes used by the relation queries, including tempo, key, instruments, and genres. We use production metadata when available; otherwise, these fields can be estimated from full-song audio using standard MIR tools for structure segmentation, source separation, beat tracking, key estimation, and instrument tagging. This allows other full-song corpora to be represented in the same schema.
The catalog uses a relational schema, with one row per excerpt in the main table and separate tables for set-valued genre and instrument fields.

\begin{table*}[t]
\centering
\caption{The nine relations: the conditions a positive tuple must satisfy, the
units each is instantiated over, and the question a candidate QA instance asks.
A pair QA instance asks the same relation of two excerpts. In the Units column,
\emph{Sect.} denotes a section excerpt, and \emph{Stem} denotes a section stem.}
\label{tab:relations}
\scriptsize
\setlength{\tabcolsep}{2.5pt}
\begin{tabular}{@{}lllp{0.40\textwidth}@{}}
\toprule
Relation & Positive conditions & Units & Question \\
\midrule
\multicolumn{4}{@{}l}{\textit{Structural}} \\
same track & \texttt{same(track), different(section)} & Sect., Stem & Which candidate audio comes from the same track as Audio A? \\
adjacent section & \texttt{same(track), section\,$+$\,1} & Sect., Stem & Which candidate excerpt immediately follows Audio A in the original track? \\
same section & \texttt{same(track), same(section), different(role)} & Stem & Which candidate audio is a different isolated component from the same recorded section as Audio A? \\
parent--child & \texttt{same(track), same(section)} & Sect.\,$\to$\,Stem & Which candidate audio is contained within Audio A? \\
\midrule
\multicolumn{4}{@{}l}{\textit{Attribute}} \\
same key & \texttt{same(key), different(track)} & Sect., Stem & Which candidate audio shares the same musical key as Audio A? \\
same tempo & \texttt{close(tempo,3), different(track)} & Sect., Stem & Which candidate audio has a similar tempo to Audio A? \\
tempo order & \texttt{greater(tempo,3), different(track)} & Sect., Stem & Which candidate audio is faster in tempo than Audio A? \\
instrument overlap & \texttt{overlap(instr,2), different(track)} & Sect. & Which candidate audio shares at least two instrument tags with Audio A? \\
genre overlap & \texttt{overlap(genre,2), different(track)} & Sect. & Which candidate audio shares at least two genre tags with Audio A? \\
\bottomrule
\end{tabular}
\end{table*}

\subsection{Relation queries}

A \emph{relation} is defined as a declarative record. It specifies the source of the anchor, correct candidate, and distractor, as well as the conditions for valid positive and hard-negative tuples. For example, a \emph{parent--child} relation pairs a section with one of its stems. Each condition applies one of eight operators to a catalog field: \texttt{same} and \texttt{different}; \texttt{close} and \texttt{far} with a tolerance; \texttt{overlap} and \texttt{no\_overlap} with a minimum; and \texttt{greater} and \texttt{less}. Table~\ref{tab:relations} lists the positive conditions and corresponding questions for each relation. For each relation, hard-negative conditions preserve selected context while violating the target relation. All excerpts in a tuple are distinct and non-silent.

A \emph{QA format} specifies how the audio evidence is presented and how the answer is expressed.
In \emph{pair QA}, a question about two excerpts is answered with yes/no or a short value.
In \emph{candidate QA}, an anchor is paired with four audio candidates and the model selects one, forming a multiple-choice question whose options are audio rather than text. Because relation and format are independent, one relation can define multiple tasks: \emph{same track}, for example, can ask either ``Do these two excerpts come from the same track?'' or ``Which candidate comes from the same track as the anchor?''

A single SQL query over the catalog generates each task tuple, its answer, and the answer position. The query has one branch per role, while a hash determines the slot of the correct candidate. One \emph{same section} instance, with identifiers written as \emph{track-section-role}, is:

\vspace{2pt}
\noindent\scriptsize
\begin{tabular}{@{}l@{\hspace{2mm}}l@{\hspace{3mm}}l@{}}
\toprule
\multicolumn{3}{@{}l}{\textbf{catalog} \textrm{(\texttt{samples})}} \\
\multicolumn{3}{@{}l}{%
\begin{tabular}{@{}l@{\hspace{1.5mm}}l@{\hspace{3mm}}l@{\hspace{1.5mm}}l@{}}
anchor & \texttt{726150-Verse-vocals} & hard-negative & \texttt{726150-Chorus-drums} \\
positive & \texttt{726150-Verse-drums} & negative & \texttt{128662-Verse-drums} \\
 & & & \texttt{1074236-Inst-drums} \\
\end{tabular}} \\
\midrule
\multicolumn{2}{@{}l}{\textbf{relation} \textrm{\emph{same section}}} & \textbf{compiled query} \\
positive & \texttt{same(track)} & \texttt{b.track\_id = a.track\_id} \\
 & \texttt{same(section)} & \texttt{b.section\_id = a.section\_id} \\
 & \texttt{different(role)} & \texttt{b.stem\_role <> a.stem\_role} \\
hard-negative & \texttt{same(track)} & \texttt{b.track\_id = a.track\_id} \\
 & \texttt{different(section)} & \texttt{b.section\_id <> a.section\_id} \\
\bottomrule
\end{tabular}
\normalsize
\vspace{2pt}

\noindent The positive example shares the anchor's section but uses a different role. The hard negative keeps the same track but comes from a different section, making it another stem from elsewhere in the same song. A set-valued condition such as \texttt{overlap(genre,2)} is implemented as a join over the corresponding side table with a minimum-count threshold.

\section{Experimental Setup}
\label{sec:setup}

\textbf{Corpora.}
We build the catalog from three full-song collections. An in-house production
set of 15,856 songs from a licensing-cleared catalog carries authored metadata: section boundaries, separated
stems, tempo, key, genre, and instrumentation. The other two are public and
carry no such annotation, and are included to show that the catalog does not
depend on production data: 69,297 songs from JamendoMaxCaps~\cite{jamendomaxcaps}, and 12,119 songs
under 600 seconds from a commercially reusable subset of the Free Music Archive~\cite{fma},
restricted to public-domain, CC-BY, and CC-BY-SA licenses at a pinned release.
Before generation, we remove thirteen JamendoMaxCaps tracks whose audio matches
evaluation items; the corpus counts above are after this exclusion. A full song is represented by its sections concatenated
in order, or by the original recording where no segmentation exists.

\textbf{Catalog construction.}
For the two public corpora every field of Section~\ref{sec:catalog} is estimated
from audio: sections from SongFormer~\cite{songformer}, stems from
SCNet~\cite{scnet}, key and tempo from Essentia~\cite{essentia}, and genre and
instrument tags from Discogs-EffNet models~\cite{discogseffnet}. Estimated fields are
stored exactly like authored ones, so a relation is written once and holds over
every corpus without change.

\textbf{Training data.}
\train{} is the instruction-tuning half of \method{}. Its two roles, Relation
MultiQA and coreference, contribute 97,272 instances each, so \train{} holds
194,544 instances with one to five audio inputs each. Relation instances are
$48\%$ section-only, $45\%$ section-stem-only, and $7\%$ mixed. Relation MultiQA
is the multi-audio data of Section~\ref{sec:method}, generated at one instance per full
song.
Coreference is assembled from a pool of single-audio full-song QA, built
by adapting Music Flamingo's synthetic-caption recipe~\cite{musicflamingo} and
the MOSS-Music-Data-Pipeline\footnote{\href{https://github.com/wx9Songs/MOSS-Music-Data-Pipeline}{\nolinkurl{github.com/wx9Songs/MOSS-Music-Data-Pipeline}}.}: an audio-language model drafts an
audio-grounded caption and a text LLM refines it against source metadata and MIR
estimates.

\textbf{Coreference.}
A multi-audio instruction carries references that mean nothing outside its
context, such as \texttt{Audio A}, so answering it requires grounding language
to the correct input index~\cite{polyaudio}. Two to four full songs from
distinct tracks are concatenated under separate labels, and the target answer
repeats each song's own full-song answer under its matching label, so a correct
answer places each description under the label of the audio it came from.

\textbf{Models and instruction format.}
We instruction-tune two backbones, Music Flamingo and
MOSS-Music~\cite{mossmusic2026}. Both are already instruction-tuned for
single-audio music understanding over full songs, the setting \train{} extends
rather than replaces. Each is fully fine-tuned for one epoch
with batch size $128$ and learning rate $1\times10^{-5}$ on two B200 GPUs.
Every instance places one \texttt{<audio>} placeholder per input on its own labeled
line and uses letters rather than indices. Pair QA uses \texttt{Audio A} and
\texttt{Audio B}; candidate QA uses \texttt{Audio A} with
\texttt{Candidate B}--\texttt{E}. Each prompt closes with an explicit
response-format instruction. Training and evaluation use
the same contract, re-serialized into each backbone's native audio interface.

\textbf{Evaluation.}
Multi-audio evaluation uses \bench{}, our held-out benchmark of candidate QA
posed over sections and section stems drawn from full songs.
It is
generated by the same procedure as \train{}'s Relation MultiQA but from MUSDB18~\cite{musdb18}, MedleyDB~\cite{medleydb}, and MoisesDB~\cite{moisesdb}, thus the two share
no corpus and its stems are authored rather than separated.
It holds 300 questions (100 per corpus) over 481 songs, of which 127 test structural and 173
musical-attribute relations.
Every item was checked by hand, and those
judged musically implausible were replaced under the same source, relation and
answer strata until all $300$ were accepted. Two further multi-audio benchmarks are reported
beside it: MUGEN: Music Analysis~\cite{mugen}, and $2{,}010$ short-answer items
from Jamendo-MT-QA~\cite{jamendomtqa}. Single-audio music understanding is
measured on MMAU-Pro: Music~\cite{mmaupro}, MMAU: Mini-Music~\cite{mmau}, MMAR:
Music~\cite{mmar} and RUL-MuChoMusic~\cite{rulmusic}, the last under its
four-fold protocol. Direct systems answer with greedy decoding, and each
component of a cascade uses its own default sampling parameters; a response that
cannot be parsed into the requested format counts as incorrect.

\nocite{qwen36}%
\begin{table*}[!t]
\centering
\caption{System-level accuracy across multi-audio reasoning and single-audio music understanding. Each cascade supplies independently generated audio captions to Qwen3.6-35B-A3B. Best and second-best scores are bold and underlined, respectively. Parse failures count as incorrect.}
\label{tab:system_comparison}
\resizebox{\textwidth}{!}{%
\begin{tabular}{lcccccccc}
\toprule
& \multicolumn{4}{c}{Multi-Audio QA} & \multicolumn{4}{c}{Single-Audio QA} \\
\cmidrule(lr){2-5}\cmidrule(lr){6-9}
System & MUGEN & Jamendo-MT-QA & \multicolumn{2}{c}{\bench{}} & MMAU-Pro & MMAU & MMAR & RUL-MuChoMusic \\
\cmidrule(lr){4-5}
 & Music Analysis & & Struct. & Attr. & Music & Mini-Music & Music & \\
\midrule
\multicolumn{9}{l}{\textit{Cascaded systems}} \\
Audio Flamingo 3~\cite{audioflamingo3} $\rightarrow$ Qwen3.6 & 0.5000 & 0.7169 & 0.4803 & 0.2428 & 0.5896 & 0.7006 & 0.4138 & \textbf{0.6828} \\
Music Flamingo $\rightarrow$ Qwen3.6 & 0.5400 & \textbf{0.8716} & 0.4803 & 0.2428 & 0.5367 & 0.6677 & 0.4039 & 0.6114 \\
MOSS-Audio~\cite{mossaudio2026} $\rightarrow$ Qwen3.6 & 0.5320 & 0.7741 & 0.4094 & 0.2254 & 0.4915 & 0.5689 & 0.4236 & 0.5078 \\
MOSS-Music $\rightarrow$ Qwen3.6 & 0.5800 & \underline{0.8692} & 0.5039 & 0.2601 & 0.4972 & 0.5778 & 0.3892 & 0.5082 \\
\midrule
\multicolumn{9}{l}{\textit{Direct QA systems, pretrained}} \\
Qwen2-Audio~\cite{qwen2audio} & 0.1960 & 0.5348 & 0.2283 & 0.2254 & 0.4337 & 0.6018 & 0.3941 & 0.4804 \\
Qwen2.5-Omni~\cite{qwen25omni} & 0.5400 & 0.7582 & 0.5354 & \underline{0.2948} & 0.6601 & 0.6826 & 0.5123 & 0.4530 \\
Audio Flamingo 3 & 0.1720 & 0.5627 & 0.2126 & 0.1676 & 0.6361 & 0.7455 & 0.4828 & \underline{0.6590} \\
MOSS-Audio & 0.3720 & 0.6433 & 0.3780 & 0.2370 & 0.6481 & 0.7635 & 0.5517 & 0.5179 \\
Music Flamingo & 0.2760 & 0.7836 & 0.2205 & 0.1734 & 0.6700 & 0.7395 & 0.4828 & 0.2207 \\
MOSS-Music & \underline{0.6240} & 0.8373 & 0.3543 & 0.2832 & \underline{0.7193} & \underline{0.7665} & \underline{0.5764} & 0.5792 \\
\cmidrule(lr){1-9}
\multicolumn{9}{l}{\textit{Direct QA systems, fine-tuned}} \\
Music Flamingo w/ \train{} & 0.5640 & 0.8478 & \textbf{0.8189} & \textbf{0.3526} & 0.6770 & 0.7455 & 0.5123 & 0.3012 \\
MOSS-Music w/ \train{} & \textbf{0.6320} & 0.8622 & \underline{0.5906} & 0.2890 & \textbf{0.7228} & \textbf{0.7844} & \textbf{0.6010} & 0.5866 \\
\bottomrule
\end{tabular}}
\end{table*}

\begin{table}[!t]
\centering
\caption{Ablation over the two multi-audio roles of \train{}; Relation MultiQA
$+$ Coreference is \train{} itself. Each average weights the columns of
Table~\ref{tab:system_comparison} equally within its group. Parentheses report
percentage-point changes from the backbone's pretrained model.}
\label{tab:ablation}
\scriptsize
\renewcommand{\arraystretch}{1.15}
\begin{tabular*}{\columnwidth}{@{\extracolsep{\fill}}lcc@{}}
\toprule
Training data & Multi-Audio Avg. & Single-Audio Avg. \\
\midrule
\multicolumn{3}{@{}l}{\textit{Music Flamingo}} \\
Pretrained & 0.3634 & 0.5282 \\
Coreference only & 0.4189 ($+5.55$) & 0.5345 ($+0.63$) \\
Relation MultiQA only & 0.5463 ($+18.29$) & 0.5538 ($+2.56$) \\
Relation MultiQA $+$ Coreference & {\boldmath\textbf{0.6458 ($+28.24$)}} & {\boldmath\textbf{0.5590 ($+3.08$)}} \\
\midrule
\multicolumn{3}{@{}l}{\textit{MOSS-Music}} \\
Pretrained & 0.5247 & 0.6603 \\
Coreference only & 0.5331 ($+0.84$) & {\boldmath\textbf{0.6772 ($+1.69$)}} \\
Relation MultiQA only & 0.5597 ($+3.50$) & 0.6640 ($+0.37$) \\
Relation MultiQA $+$ Coreference & {\boldmath\textbf{0.5935 ($+6.88$)}} & 0.6737 ($+1.34$) \\
\bottomrule
\end{tabular*}
\end{table}

\section{Results}
\label{sec:results}

\subsection{System-level comparison}

Table~\ref{tab:system_comparison} compares direct audio-QA models with cascaded
systems, in which an audio-language model captions every input independently and
Qwen3.6-35B-A3B~\cite{qwen36} reasons over the resulting captions.

Tuning on \train{} improves every multi-audio benchmark for both backbones. Music
Flamingo gains $28.8$ points on MUGEN: Music Analysis, $6.4$ on Jamendo-MT-QA,
$59.8$ on \bench{}'s structural relations and $17.9$ on its attribute relations;
MOSS-Music gains $0.8$, $2.5$, $23.6$ and $0.6$. The gain is larger for backbones with lower pretrained performance: MOSS-Music is already the strongest pretrained direct
system on MUGEN: Music Analysis ($0.6240$) and moves least there, while Music
Flamingo starts at $0.2760$ and gains most.
Single-audio music understanding is also retained for both backbones and improves
on every benchmark, with the largest gain coming from Music Flamingo on
RUL-MuChoMusic ($+8.1$).

What separates the benchmarks is how far a cascade gets. On Jamendo-MT-QA the
best system in the table is a cascade, $0.8716$ against $0.8622$ for the best
tuned model, which matches Jamendo-MT-QA's own report that caption-based
baselines remain competitive with multi-audio systems~\cite{jamendomtqa}. On
\bench{}'s structural relations the ordering reverses and the margin is wide: the
best tuned model reaches $0.8189$ against $0.5039$ for the best cascade, and
tuned Music Flamingo exceeds its own cascade by $33.9$ points.
Comparisons
between independently collected tracks are largely answerable from per-track
descriptions; relations among excerpts that share a source are not.

Among pretrained systems the strongest on \bench{} is Qwen2.5-Omni, which is not
music-specialized: $0.5354$ on the structural relations against $0.3543$ for
MOSS-Music and $0.2205$ for Music Flamingo, and $0.2948$ on the attribute
relations, second in that column only to the tuned Music Flamingo.  These results suggest that native support for multiple audio inputs may matter more here than music-specific pretraining.

\subsection{Relation-level analysis}

Table~\ref{tab:system_comparison} splits \bench{} into the relations whose answer
the catalog records as provenance and those whose fields are estimated by MIR
components, and the two groups behave differently. Tuning moves the structural
relations by $59.8$ points for Music Flamingo and $23.6$ for MOSS-Music, and
Music Flamingo reaches $0.8189$, the highest score in that column. The attribute
relations top out at $0.3526$ against a chance of $0.25$, and most systems in the
table sit at or below chance on them.

The cascades give the same reading from the other side. With a strong captioner
and a text model they reach $0.4094$--$0.5039$ on the structural relations but
only $0.2254$--$0.2601$ on the attribute relations. Why the attribute relations stay
hard is open. They depend on intentionally small differences, e.g., three beats
per minute, one key label, or two shared tags. Such fine-grained differences may
be lost both in captions and in the MIR estimates used to populate the catalog.

\subsection{Training-mixture analysis}

Table~\ref{tab:ablation} separates the two multi-audio roles of \train{}.
Coreference alone moves multi-audio accuracy less, $+5.55$ points for Music
Flamingo and $+0.84$ for MOSS-Music, against $+18.29$ and $+3.50$ for Relation
MultiQA alone.
The two roles teach different things: holding several inputs
apart, and knowing what relates them.

Training on both beats either alone, and by more than the two add up to:
$+28.24$ against $23.84$ for Music Flamingo, $+6.88$ against $4.34$ for
MOSS-Music. Single-audio accuracy stays above the pretrained score in every
condition.

\section{Conclusion}

\method{} uses song--section--stem provenance to generate relation-first
multi-audio QA and instruction-tuning data.
On \bench{}'s structural relations, both tuned models outperform their
corresponding caption cascades, whereas a cascade achieves the highest score on
Jamendo-MT-QA.
Combining Relation MultiQA with coreference yields the highest multi-audio average
for both backbones while keeping their single-audio averages above pretrained
levels.

\clearpage
\section{Acknowledgments}

This work was supported by the Korea Creative Content Agency (KOCCA) and the Artificial Intelligence Industry Cluster Agency (AICA).

\bibliographystyle{IEEEbib}
\bibliography{strings,refs}

\end{document}